 \documentstyle[11pt,aaspp4, flushrt]{article}
\begin{document}
\title{New Results for Two Optically Faint Low Mass X-Ray Binary Systems}
\author{Stefanie Wachter\altaffilmark{1}}  
\affil{Department of Astronomy, University of Washington, Box 351580,
 Seattle, WA 98195-1580; wachter@astro.washington.edu}
\altaffiltext{1}{Visiting Astronomer, Cerro Tololo Interamerican Observatory,
National Optical Astronomy Observatories, operated by AURA, Inc. under 
cooperative agreement with the NSF.}
\authoremail{wachter@astro.washington.edu}

\begin{abstract}

We present optical photometry of the low mass X-ray binary systems
GX~349+2 (=X1702-363) and Ser~X-1 (=X1837+049). 
Extensive $VRI$ photometry of the faint optical counterpart ($V=18.4$)
to GX~349+2
reveals a period of $22.5 \pm 0.1$~h and half-amplitude 0.2~mag. 
This result confirms and extends
our previously 
reported 22~h period (Wachter \& Margon 1996).  
No color change is detected over the orbit, although the limits are 
modest.
We also report the discovery of two new variable stars in the field of 
GX~349+2, including a probable W~UMa system. 

Ser~X-1 is one of the most intense persistent X-ray burst
sources known. It is also one of only three burst systems
for which simultaneous optical and X-ray bursts have been observed.
The faint blue optical counterpart MM~Ser ($B\sim19.2$) has 
long been known to have a companion 2.1\arcsec~distant. 
Our images indicate 
that MM~Ser is
itself a further superposition of two stars, separated by only 1\arcsec.
At the very least, the ratio of inferred burst to quiescent optical flux
is affected by the discovery of this additional component. In the worst
case, the wrong object may have previously been assumed as the optical 
counterpart.

\end{abstract}
 

\section{Introduction}

Little is known about the faint optical counterpart ($V = 18.4$) of the
bright bulge Z-source GX~349+2.
The one published spectrum of the counterpart
is of both low
resolution and quality (Penninx \& Augusteijn 1991) and only shows
a strong H$\alpha$
emission line.
From the first photometric study of the system we reported the detection
of a possible $21.85\pm 0.13$~h period
(Wachter \& Margon 1996, hereafter Paper I). At the same time,
Southwell et al. (1996) suggested a period of 14~d
for GX~349+2 from an analysis
of the H${\alpha}$ emission line velocities.
In order to resolve this ambiguity and better determine the orbital period we
undertook a second monitoring campaign of the system.
 
Ser~X-1 is one of the most intense persistent X-ray burst
sources known.
It was discovered in X-rays in 1965 (Friedman, Byram \& Chubb 1967)
and to be a burst source in 1975 (Swank et al. 1976).
A faint blue star of $B\sim18.5$ was suggested as the optical counterpart by
Davidsen (1975). Thorstensen, Charles \& Bowyer (1980), hereafter TCB,
 resolved the Davidsen candidate
(star D) into two stars (DN and DS) separated by 2.1\arcsec. The southern
component, now known as MM~Ser, which TCB
proposed as the
X-ray source counterpart, is the bluer of the two stars and possibly exhibits
a weak HeII $\lambda$4686 emission line, although the published spectra are 
quite noisy. 
There has been no detailed photometric study of Ser~X-1. 
 
The optical identification of Ser~X-1 is especially important as it is one of
only three low mass X-ray binaries (LMXBs) 
(the others being X1636-536 and X1735-444) for which
simultaneous optical and X-ray bursts have been observed (Hackwell et al.
1979). The optical burst is thought to be due to reprocessing of the X-ray
burst in matter in the close vicinity of the neutron star, either the
accretion disk or the atmosphere of the companion.

\section{Observations}

$VRI$ CCD photometry of GX~349+2 and Ser~X-1 was performed with the CTIO  
0.9~m telescope
from 1996 July 10 UT to July 15 UT. 
A Tek 2048 CCD was used at an image scale of 0.40\arcsec\ pixel$^{-1}$.
A summary of the observations of GX~349+2 is 
provided in Table~\ref{t1}. $N_V$, $N_R$, and $N_I$ refer to the number of 
$V$, $R$ and $I$ frames obtained during each night. Exposure times for GX~349+2 
were typically 600~s ($V$), 400~s ($R$), 300~s ($I$) and 600~s for Ser~X-1.  
Overscan and bias corrections were made for each CCD image with the 
task {\it quadproc} at CTIO to deal with the 4 amplifier readout.  
The data were flat-fielded in the standard 
manner with IRAF.

Due to the crowded field
photometry was performed by point spread function (PSF) fitting with DAOPHOT~II
\markcite{S}(Stetson 1993). 
The instrumental magnitudes were transformed to the standard system through
observations of four Landolt standard star fields (providing 20 standard stars)
on one of the nights  
(\markcite{L}Landolt 1992).
Using the standardized photometry from the photometric night, we selected
nine local standard stars in a 3.5\arcmin$\times$3.5\arcmin~field around 
GX~349+2 and Ser~X-1.
Stars were chosen as local standards if they were uncrowded and 
relatively bright.
We then performed relative photometry of GX~349+2 
and derived standardized magnitudes  
with respect to each local standard, including corrections for color terms. 
Finally, we computed the 
mean and standard deviation of the
nine estimates, using a 3$\sigma$ clipped mean to remove outliers. 
The magnitudes for GX~349+2 and two comparison stars are listed in 
Table~\ref{t2}; the magnitudes for Ser~X-1 and comparison stars 
in Table~\ref{t3}.  

The external 1$\sigma$ error (from the transformation
to the standard system) is about 0.03~mag.
Intrinsic errors were calculated by computing the average rms scatter in the 
light curves of all non-variable 
stars in the field. In the magnitude range $R \leq 18.0$, this intrinsic
1$\sigma$ error is 0.01~mag, and in the range $18.0 \leq R \leq 19.0$,
0.01--0.025~mag.

\section{GX~349+2}

\begin{figure}[h]
\epsscale{0.6}
\plotone{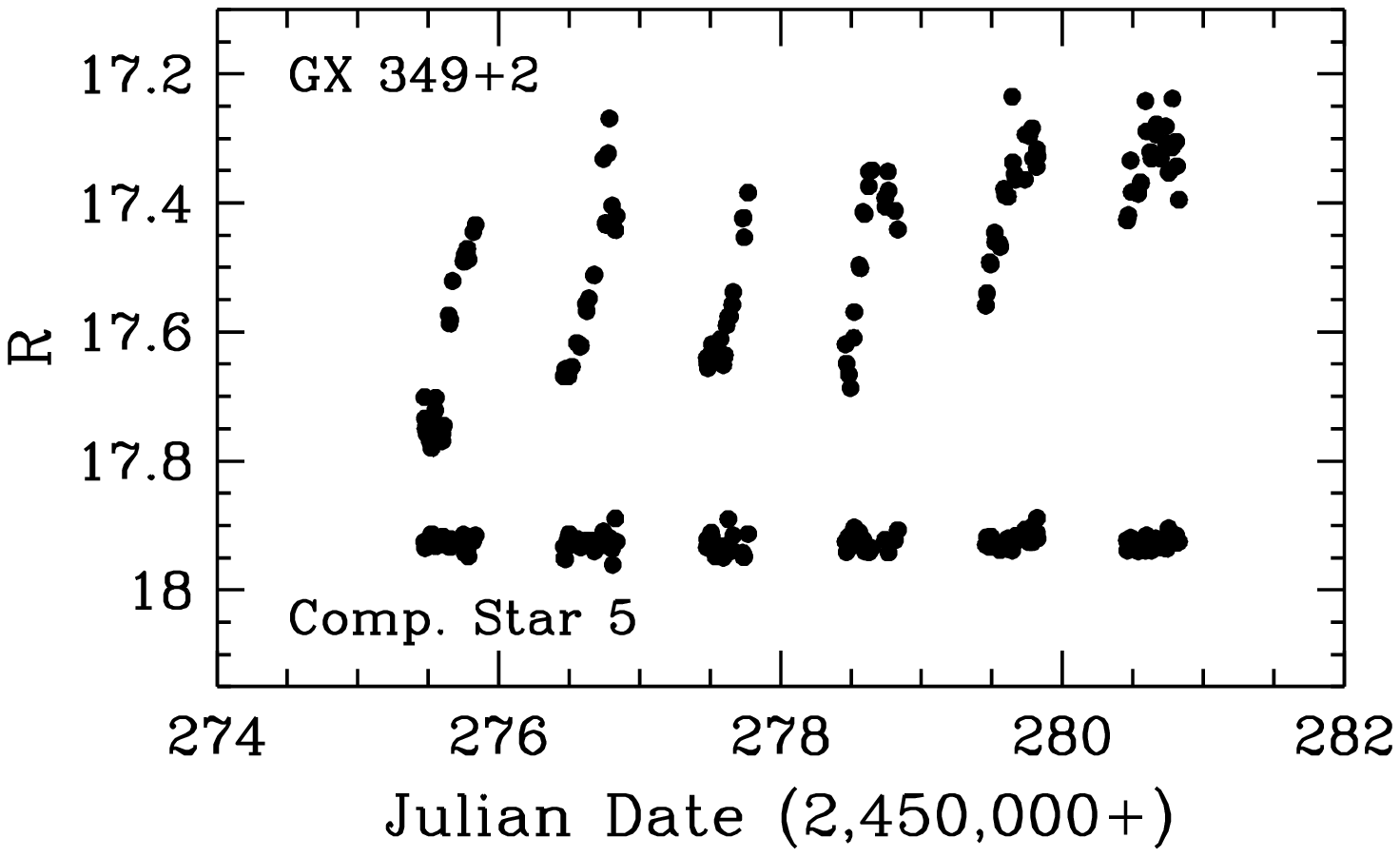}
\caption[]{The light curve of GX~349+2 and a comparison
star of similar brightness. The magnitudes of comparison star 5
have been shifted to fainter magnitudes by 0.2~mag for clarity of display.
The 1$\sigma$ error is 0.01~mag,
obtained from the
rms scatter in the light curves of the comparison stars.
 \label{f-lc}}
\end{figure}

The $R$ band light curves for
GX~349+2 and a slightly fainter comparison star (star 5)
are displayed in Figure~\ref{f-lc}. For the star designations refer
to the finding chart in Paper I.  
The magnitude of the comparison
light curve has been shifted to fainter magnitudes by 0.2~mag to separate
the two curves for display purposes. The pronounced variability of GX~349+2
is clearly evident on each night. 

\begin{figure}
\epsscale{0.6}
\plotone{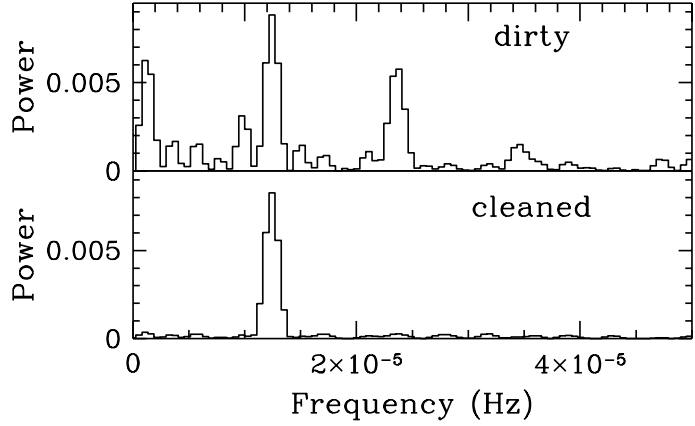}
\caption[]{Power spectrum analysis with the CLEAN algorithm. {\it Top panel:}
the dirty spectrum, i.e., the convolution of the periodicities
in the data with the sampling window. {\it Bottom panel:} the cleaned spectrum
after 10 iterations with a gain of 0.2, displaying the strong peak at 22.4~h.
\label{f-power}}
\end{figure}

The data were searched for periodic
behavior with a power spectrum analysis using the CLEAN algorithm 
(Roberts
et al. 1987).  
After 10 iterations of CLEAN with a gain of 0.2, the cleaned
power spectrum exhibits a single strong peak at $22.4 \pm 0.5$~h (Figure~\ref{f-power}).
We searched the period range around this peak according to the procedure 
described in Paper I. The resulting best period and $1\sigma$ error 
is $22.5 \pm 0.1$~h, 
confirming and refining our previous period estimate.  A period search 
carried out with the phase dispersion minimization algorithm {\it PDM}
in IRAF arrives at the same period. 
The 1~d alias of our 22.5~h period is 15.0~days, which is close to the 14~d 
period reported
by Southwell et al. (1996). Their period of $13.94\pm 0.1$~day 
corresponds to the 1~day alias of $22.39\pm 0.01$~h, and this may explain 
their reported periodicity. 

\begin{figure}
\epsscale{0.6}
\plotone{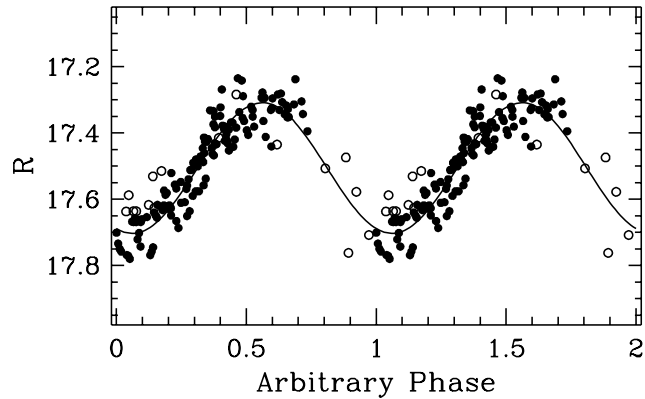}
\caption[]{The light curve
for our photometric data
of GX~349+2, folded on the best
period of 22.5~h. Also plotted is the best-fitting sine wave.
Two cycles are shown for clarity. Filled circles mark the July 1996 data,
open circles represent additional data obtained about three weeks earlier.
\label{f-fold}}
\end{figure}

In addition to the July 1996 data, supplementary data of GX~349+2 were obtained by
{\mbox E. W. Deutsch} and
{\mbox B. Margon} about three weeks 
earlier (June 1996). When the above period analysis is repeated with the combined data set, 
the formal best period and $1\sigma$ error are $22.53\pm 0.02$~h. However, the 
June data are very sparsely sampled (see Table~\ref{t1}) and the source displays
considerable intrinsic scatter in the folded lightcurve (Figure~\ref{f-fold}).
In fact, there is no obvious preferred period when comparing the data folded on 
periods between 22.4 -- 22.6~h, so that the precision in the above period seems 
overly optimistic. Consequently, we adopt 22.5~h as our best period estimate for the
following discussion. 

Figure~\ref{f-fold} shows our photometric data folded on the best 
period of 22.5~h and fitted with a sine wave. Filled circles mark the July 1996 data, 
open circles
represent the June 1996 data. 
Two cycles are shown 
for clarity. The sine fit gives a half-amplitude of 0.20~mag for the 
variations, comparable to that reported in Paper I. Erratic flickering is 
evident in the light curve, as is 
often observed in LMXBs.

The refinement of the period of course does not alter the conclusions 
reached in 
Paper I with respect to the mass function or the components of the binary. 
If we assume our 22.5~h period to be the orbital period of the system 
(although it could also be half the orbital period) and if the 
Southwell et al. (1996) spectroscopic observations define the true 
radial velocity 
semi-amplitude of the system ($K\approx70$~km~s$^{-1}$),  
then the implied mass function is 
$f(M) = 0.033$~M$_{\odot}$ (essentially identical to the estimate in  
Paper I). 
Of course one has to be cautious when deriving mass functions from emission 
line velocity curves for these interacting systems. The emission lines are
commonly thought to originate in the accretion disk. If this emission is asymmetric,
then the observed radial velocity along a given line of sight will combine
components of the orbital velocity of the system and the rotational velocity 
around the central object of the emitting material in the disk.
GX~349+2 appears to be very similar to Sco~X-1,  
where the period and mass function are 18.9~h and  
$f(M) = 0.016$~M$_{\odot}$ (Cowley \& Crampton 1975), 
respectively.  
Considerable similarity in the X-ray properties of Sco~X-1 and
GX~349+2 had been noted for some time (Schulz, Hasinger, \& Tr\"umper 1989),
and this has recently been extended in a more detailed study 
{\mbox (Kuulkers et al. 1997)}.

\begin{figure}
\epsscale{0.6}
\plotone{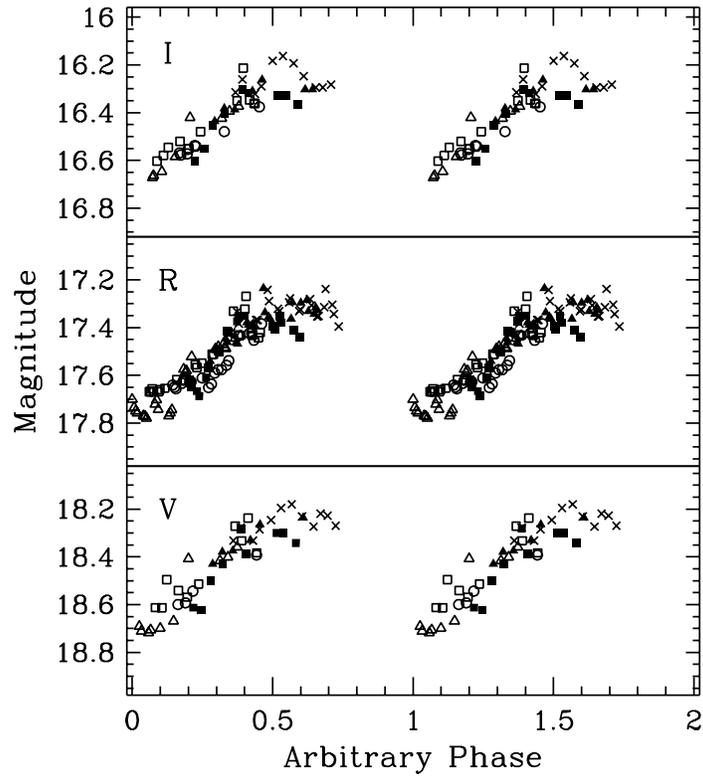}
\caption[]{The light curves of GX~349+2 in the $V$, $R$ and
$I$ bands folded on the best period of 22.5~h. The different symbols indicate
the different observing nights:
night~1 -- open triangles, night~2 -- open squares, night~3 -- open circles,
night~4 -- filled squares, night~5 -- filled triangles,
night~6 -- crosses.
\label{f-vri}}
\end{figure}

\begin{figure}
\epsscale{0.6}
\plotone{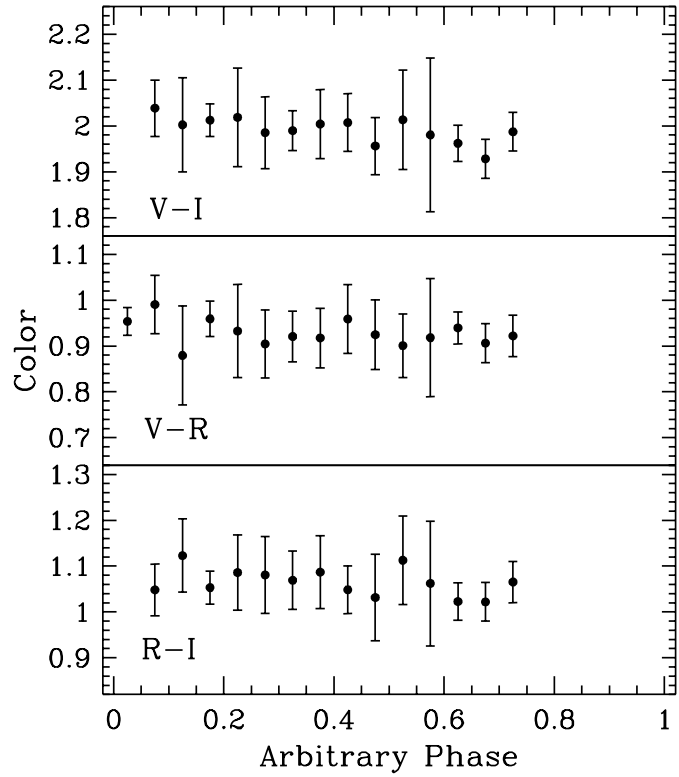}
\caption[]{The variation of color of GX~349+2 over
orbital phase. The error bars were calculated by adding in quadrature the
errors in each band resulting from binning the data. There is no obvious
color change over the orbital period.
\label{f-color}}
\end{figure}

We obtained the photometric data in several filters in order to investigate 
any color changes of the system across the orbit as might be expected 
from X-ray heating. Figure~\ref{f-vri}
shows our data in the different bands folded on the best period of 22.5~h.
We derived the colors ($V-R$, $R-I$ and $V-I$)
by first binning the light curve for each band into phase bins of 0.05.
We calculated the mean magnitude and standard deviation in each phase 
bin for each filter; the appropriate magnitudes were then
subtracted. The errors in each phase bin were estimated by adding in 
quadrature the magnitude errors of the two filters. 
For bins with only one 
measurement, the error in the photometry was assigned. The result is 
displayed in Figure~\ref{f-color}.  

Within the errors there is no obvious change in color across
the orbit. The error bars are fairly large, partly due to flickering and partly
due to the limited number of observations in each phase bin. 
Unfortunately, the colors considered
here ($VRI$) are not ideal to detect color variations. 
Even for systems that show significant color changes due to strong 
X-ray heating in $U-B$ and $B-V$ (e.g. Her~X-1/HZ~Her),
almost no variability is  
visible in $V-R$ (Kilyachkov et al. 1994). Due to the fairly large amount of 
extinction towards GX~349+2 it will be very difficult to obtain accurate 
$B$ photometry for the system with a small telescope.

We have measured equatorial coordinates for GX~349+2 and several other 
objects discussed in this paper by transferring
the astrometric information contained in a digitized sky survey (DSS) reference
frame to one of our CCD images. This was accomplished
with software written in interactive data language (IDL) by E. W. Deutsch
or available in the IDL Astronomy User's Library (Landsman 1993).
The DSS astrometric reference frame is transferred to the CCD Image with a
$1\sigma$ error of 0.06\arcsec. However, there may be some systematic
offset ($\sigma\sim0.5\arcsec$) from frames based on other reference
catalogs (Russell et al. 1990).
The results are listed in 
Table~\ref{t2}. The previously published optical position for GX~349+2 
was accurate 
only to 1\arcsec~in each coordinate (Cooke \& Ponman 1991).
Our improved position agrees extremely
well with the radio position given by Cooke \& Ponman (1991), which is 
also quoted in Table~\ref{t2}.

\subsection{New Variables}

\begin{figure}
\plotone{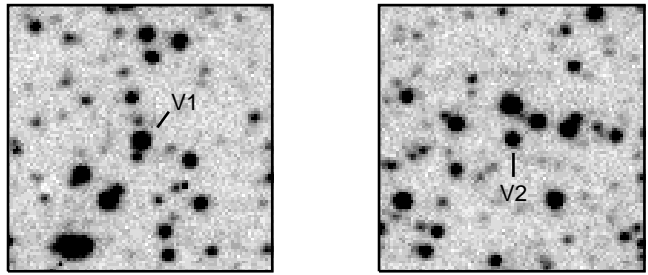}
\caption[]{$R$ band finding charts for the newly discovered
variable stars V1 and  V2. North is up and east to the left. The field size
is about 40\arcsec$\times$40\arcsec. Coordinates are given in Table~\ref{t2}.
\label{f-vfc}}
\end{figure}

We searched the 3.5\arcmin$\times$3.5\arcmin~field around GX~349+2 for 
new variable stars. For this
purpose, we computed standardized $R$ magnitudes and the rms of these 
magnitudes over the observing run for all stars in the field.
Stars with an uncharacteristically large rms in the $R$ magnitude versus rms 
diagram were identified as candidate variables. We then inspected the 
light curves of these candidates to exclude spurious detections. 
This resulted in the discovery of two new variables, designated V1 and V2,
which were not found in a search of the SIMBAD database.
Figure~\ref{f-vfc} shows
an $R$ band finding chart for each variable, and coordinates appear in 
Table~\ref{t2}. 

The magnitudes of these new variables were standardized analogous to the 
procedure outlined above for GX~349+2 and are listed in Table~\ref{t2}. 
We searched the data for periods using 
the phase dispersion minimization task PDM in IRAF. 
Our best estimate for the periods are
P$_{V1} = 38.0 \pm 3.3$~h  and
P$_{V2} = 9.34 \pm 0.03$~h. The errors were estimated by folding the data 
on a series of nearby periods and judging at what point the quality of 
the light curve is significantly compromised. 

\begin{figure}
\plotone{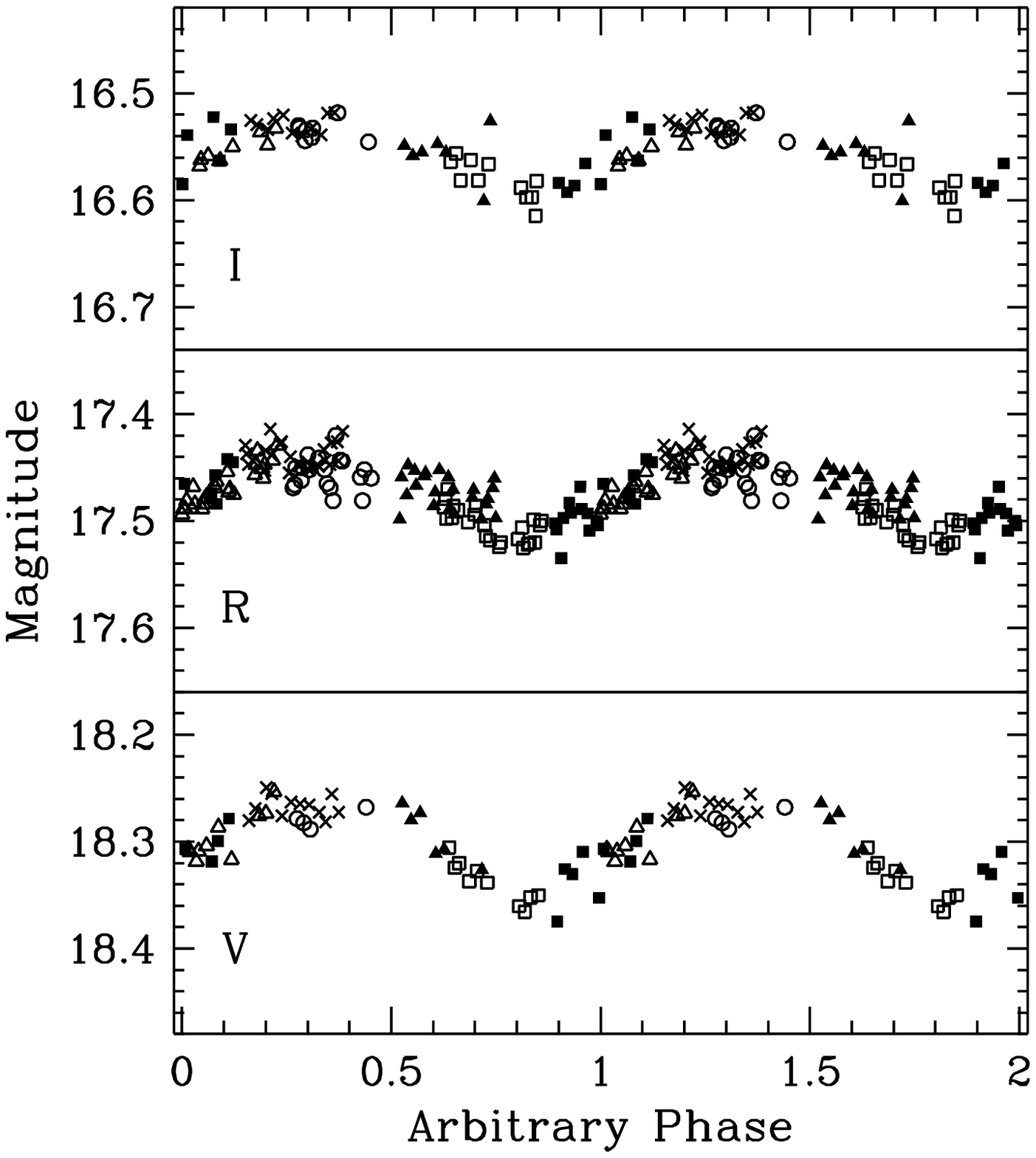}
\caption[]{The $VRI$ light curves for the newly discovered
variable V1. The data are folded on a period of 38.0~h.
Two cycles are shown for clarity.
Symbols are as described for Figure~\ref{f-vri}.
\label{f-v1}}
\end{figure}

\begin{figure}
\plotone{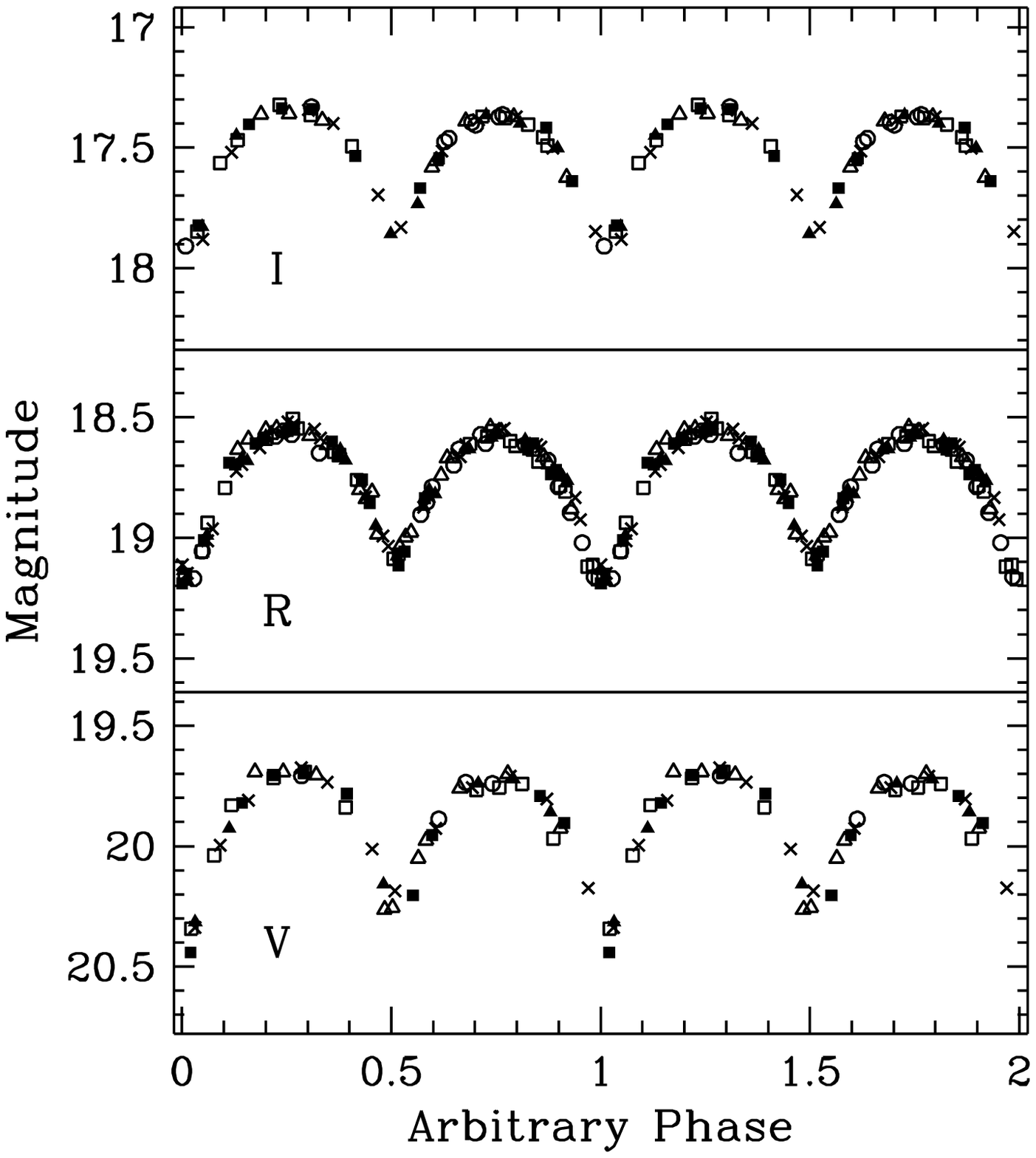}
\caption[]{The $VRI$ light curves for the newly discovered
variable V2. The data are folded on a period of 9.34~h.
Two cycles are shown for clarity.
Symbols are as described for Figure~\ref{f-vri}.
\label{f-v2}}
\end{figure}

Figures~\ref{f-v1} and \ref{f-v2} show the $VRI$ light curves folded on the
best periods for V1 and V2, respectively. 
The light curve of V2 suggests that it is a W~Ursae Majoris star, i.e., an 
eclipsing contact binary. Our best period estimate is also typical for 
W~UMa stars.   
Without any further knowledge about what causes the brightness variation in V1
it is impossible to decide whether our period  is the orbital or half the 
orbital period, i.e., whether the light curve is single or double peaked.

 \begin{table}
 \dummytable\label{t1}
 \end{table}

\begin{table}
\dummytable\label{t2}
\end{table}

\begin{table}
\dummytable\label{t3}
\end{table}

\section{Ser~X-1}

Although there is a wealth of information on the X-ray properties of 
Ser~X-1, very little is known about the optical counterpart, MM~Ser. 
Since the study of TCB, the only recently published optical
observations are two spectra of MM~Ser. 
The spectrum by Cowley, Hutchings \& Crampton (1988)
is very noisy and its only feature is possibly He~II~$\lambda4686$. A spectrum 
by Shabaz et al. (1996)\footnote{Note that the separation between DN and
DS is erroneously given as 12\arcsec~in that reference.} only shows 
an absorption feature 
at $\lambda5900$
which might be associated with a 
G star secondary. These authors determine the best fit to the 
spectrum to be that
of a G5V star with E($B-V$)=0.41$\pm0.01$. The absence of H$\beta$
absorption in the spectra of Cowley, Hutchings \& Crampton (1988) and 
Shabaz et al. (1996) seems to indicate that the detection of  
H$\beta$ absorption by TCB is due to contamination 
by the northern component.  

There has been no published photometric study of MM~Ser, and no optical 
variability (other 
than the simultaneous optical and X-ray burst) has been reported.  
TCB find no evidence for variability with a conservative limit of 
$\pm0.2$~mag (on six plates with $\sim$30~min time resolution).
In order to search for variability and possibly determine the period of the 
system we obtained several images of MM~Ser in July 1996. 

\begin{figure}
\epsscale{1.0}
\plotone{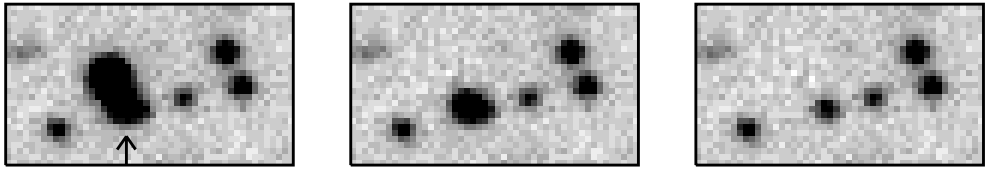}
\caption[]{{\it Left}: a portion of an $R$ band image of
the field of
Ser~X-1. North is up, and east to
the left. The arrow indicates the
widely-accepted optical counterpart of the X-ray burster, MM~Ser, the south
component of a $2''$ pair.
{\it Center}: Point-spread function (PSF) fitting is
used to remove the northern, unrelated component of the pair. Note the
east-west asymmetry of the remainder, as compared with nearby stars. {\it
Right}: another PSF fit and subtraction reveals that MM~Ser is itself
composite, consisting of
two components separated by about 1\arcsec. Although this
previously unsuspected
star uncovered by this technique is well-detected above the sky in this
frame, there are multiple, poorly constrained free parameters in the PSF
subtraction, so this resulting frame is in no way of photometric quality.
\label{f-serx1}}
\end{figure}

Due to the close companion, photometry was again performed by PSF fitting
with DAOPHOT II. Our imaging reveals that  
MM~Ser appears to be 
itself a further superposition of two stars. The leftmost image in
Figure~\ref{f-serx1}  
shows a small portion of an $R$ band image of the field of Ser~X-1.
Notice the clear separation of the components
DN and DS. In the central image,  
PSF fitting is
used to remove the northern, unrelated component of the pair. Note the
east-west asymmetry of the remainder, as compared with nearby stars. 
Another PSF fit and subtraction reveals that MM~Ser is itself
composite, consisting of
two components. 
In order to determine the separation of these components, we calculated the
mean separation of all three stars from each other from the images
with the best seeing conditions, using the positions provided as part of the 
PSF fitting by DAOPHOT II.
The separation between DN and the eastern component of DS (DSe)
is $2.07 \pm 0.02$\arcsec, and the separation between the eastern and 
western (DSw) components of
DS is $1.03 \pm 0.09$\arcsec.  

The magnitudes for the different components of star D and for
several comparison  stars are listed in Table~\ref{t3}. The nomenclature is
that of TCB. The $V$~magnitudes for the comparison
stars listed in TCB agree very well with our 
photometry. However, the comparison star $R$ magnitudes differ 
by several tenths of 
a magnitude. The reason for this difference is not clear.
TCB do not give 
$V$ magnitudes for 
DN or DS, and only list a range of 17.9--18.9 in $R$ for DS. 

\begin{figure}
\epsscale{0.6}
\plotone{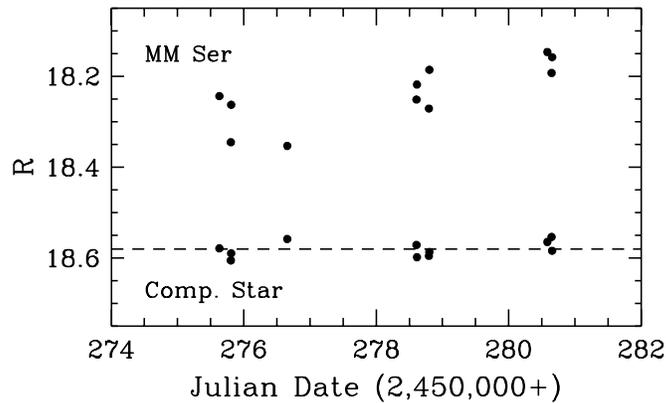}
\caption[]{The light curve of MM~Ser and a comparison
star of similar brightness. The magnitudes of the comparison star
have been shifted to fainter magnitudes by 0.25~mag for clarity of display.
\label{f-slc}}
\end{figure}

Although DS can clearly be separated into two components with PSF fitting, 
the variable PSF parameters make any photometry of the individual 
components DSe and DSw very uncertain. 
Therefore, we searched for variability in MM~Ser by removing all stars  
in its vicinity (including component DN) and subsequently performing aperture  
photometry on DS, now known to be the sum of two stars.   
The resulting photometry of DS from all $R$ frames is displayed in 
Figure~\ref{f-slc}. Also shown is a comparison star of the same brightness
which has been shifted by 0.25~mag to fainter magnitudes to separate the  
light curves. MM~Ser clearly shows variability of about 0.2~mag over   
several days and possibly variability during each night, although we cannot
state which of the two close objects is responsible. We stress that 
this conclusion 
of variability in one of the two components of MM~Ser (DSe and DSw) is 
from aperture photometry of the sum of these near-blended images, and thus
does not rely on the result of an uncertain subtraction.  

According to our photometry, the two components of DS (DSe and DSw) differ
by $\sim1$~mag in $R$.
DSe appears to be the 
bluer object, so one might speculate that this component is in fact the 
optical counterpart.  
The close companion
may severely influence spectral and radial velocity investigations of
the system. For example, it is possible that the H$\beta$ absorption observed
by TCB
originated from one of the superpositions instead of the optical counterpart
(as already mentioned above),
which could result in misleading mass estimates for the system,
as may also be true for the same reason for the X-ray binary CAL~87
(Deutsch et al. 1996). The identification of the correct optical counterpart
is also especially important for the analysis of the burst behavior, which
requires the knowledge of the correct quiescent flux of the system.
Further, accurate multicolor photometry in very good seeing conditions
might be capable of clarifying this situation. 

\acknowledgments

We thank Eric Deutsch and Bruce Margon for obtaining 
some of the observations of GX~349+2, and     
Bruce Margon
for reading a draft of this paper and 
providing helpful comments.  This research was in part supported by 
NASA grant NAG5-1630, and  
has made use of the Simbad 
database, operated at CDS, Strasbourg, France.

\newpage

\newpage

\begin{figure}
\epsscale{0.8}
\plotone{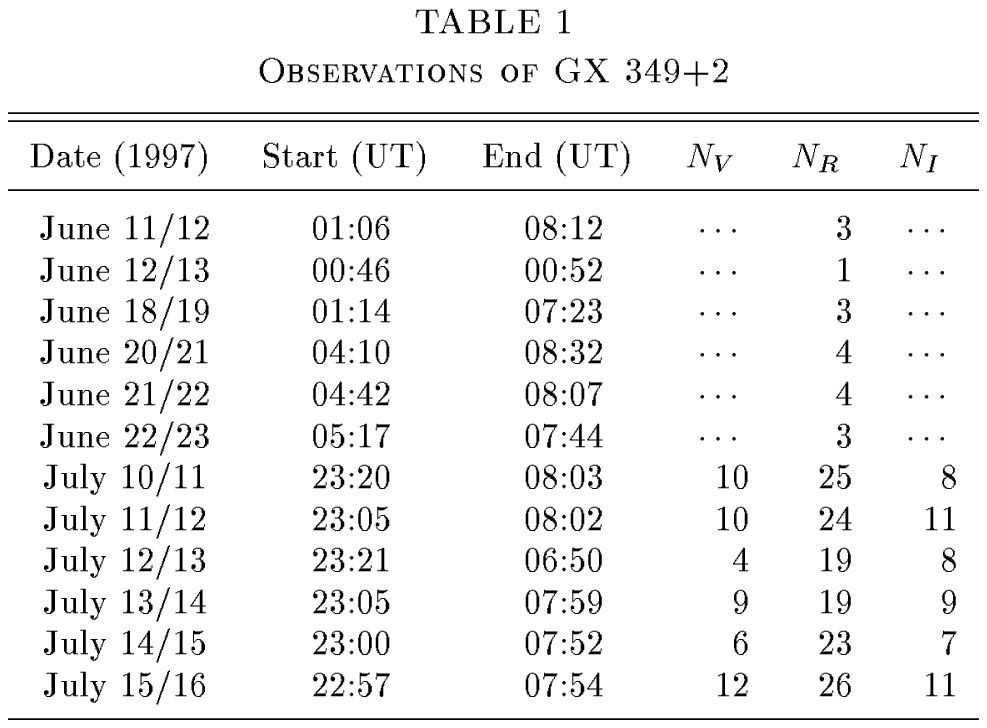}
\end{figure}

\begin{figure} 
\epsscale{1.0}
\plotone{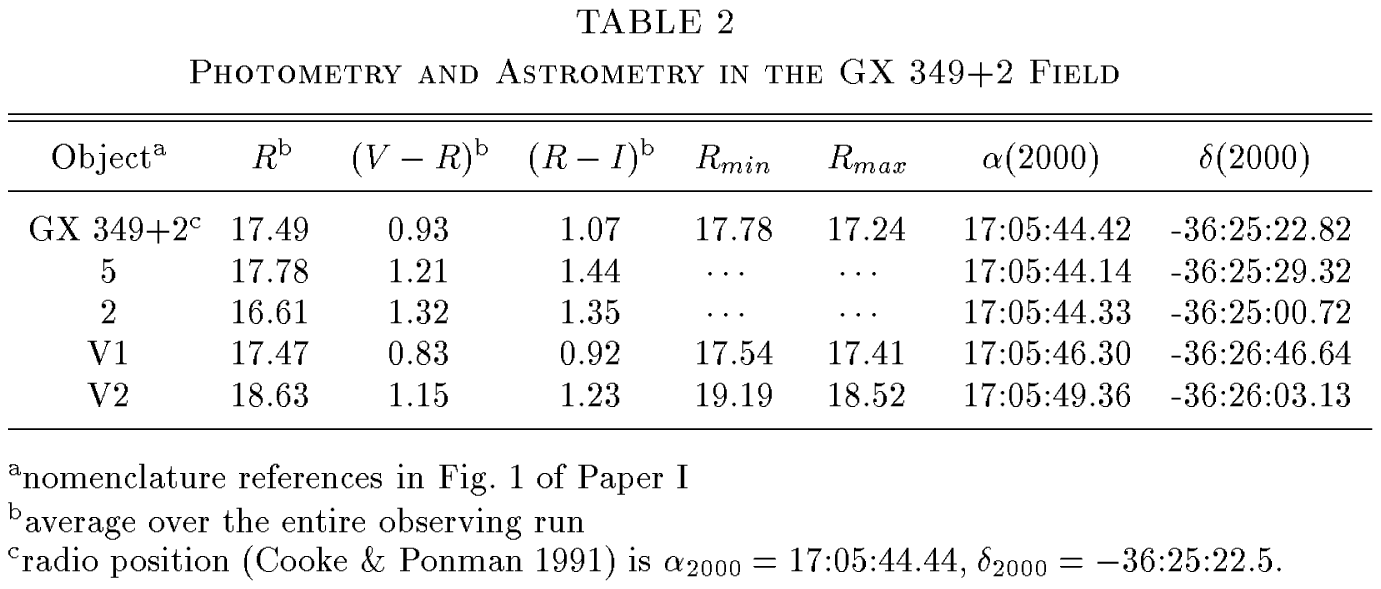} 
\end{figure} 
 
\begin{figure} 
\epsscale{0.6} 
\plotone{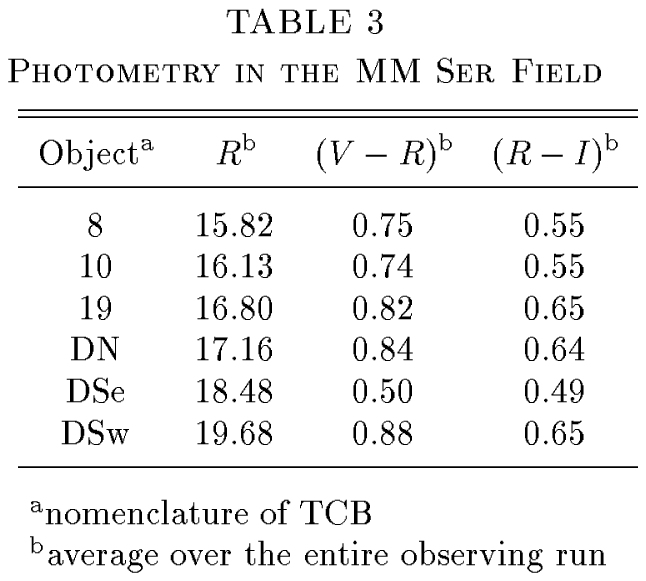} 
\end{figure}

\end{document}